\documentclass[twocolumn,aps,preprintnumbers,amsmath,amssymb]{revtex4}
\usepackage{graphicx}				
\usepackage{dcolumn}				
\usepackage{bm}					

\begin{document}
\title{Effective Rheology of Bi-Viscous Non-Newtonian Fluids in Porous Media}
\author{Laurent Talon}
\email{talon@fast.u-psud.fr}
\affiliation{Laboratoire FAST, Universit{\'e} Paris-Sud, UPMC, CNRS, Université Paris-Saclay,  F-91405, Orsay, France.}
\author{Alex Hansen}
\email{alex.hansen@ntnu.no}
\affiliation{PoreLab, Department of Physics, Norwegian University of Science and 
Technology, N--7491 Trondheim, Norway}
\date{\today}
\begin{abstract}
We model the flow of a bi-viscous non-Newtonian fluid in a porous medium by a square lattice where the links obey a piece-wise
linear constitutive equation.  We find numerically that the flow regime where the network transitions from all links behaving 
according to the first linear part of the constitutive equation to all links behaving according to the second linear part of
the constitutive equation, is characterized by a critical point.  We measure two critical exponents associated with this critical
point, one of the being the correlation length exponent.  We find that both critical exponents depend on the parameters of the
model.   
\end{abstract}
\maketitle

\section{Introduction}
\label{intro}

The behavior of complex fluids when being inside a porous medium may be very different from that when they are not.  This is a problem encountered in many biological or industrial applications ranging from impregnation of fibrous materials to immiscible multi-phase flow in porous media. Among the different types of non-Newtonian fluids, many undergo behavioral changes depending on the stress or strain applied.
One can mention the Carreau rheology  which is Newtonian at low shear rate but behaves as a power law fluid above a certain shear rate 
\cite{c72}. Other examples are yield stress fluid that responds like a solid below a critical yield threshold. Above, the materials 
behave like a power law fluid \cite{hb26}. At the mesoscopic level, this rheological approach can also be extended to other situations.
For example, inertial effects can be described as a rheological change from a Newtonian fluid to a power law (quadratic or cubic) for a certain large Reynolds number \cite{w96}. Another possible extension is the displacement immiscible fluids in porous media.
In this case, the fluids may each be Newtonian. However, the interfacial tension between them, 
makes them effectively behave in a non-Newtonian way inside the porous medium  \cite{sh12}.  
Indeed, a minimum amount of stress is then required for a non-wetting phase to invade small pore throats.

Non-Newtonian fluids are notoriously difficult to treat analytically and computationally.  When in addition the flow is constrained by the very complex boundary conditions  of the porous medium, the effective rheology of the fluid flow is not well understood. 
This might for example be seen in the fact that the leading theory for describing immiscible multi-phase flow in porous media is still the relative permeability theory dating from 1936 \cite{wb36} a theory which has evident faults.

The purpose of this manuscript is to investigate the coupling between the heterogeneities of the medium and a  rheology with a change of behavior. We will study a very simple model called a bi-viscous fluid, where the fluid is Newtonian but with a change of viscosity at one particular shear rate (or shear stress) \cite{rhg87,hrh90}.
The second viscosity might be lower (shear thinning) or higher (shear thickening).
As we will see, the coupling between the disorder and such a simple rheological model is enough to generate a rich problem.

\begin{figure}
\includegraphics[width=\hsize]{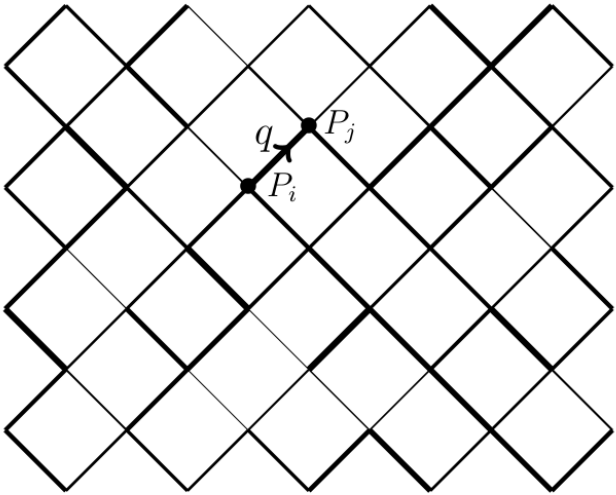}
\caption{\label{fig1} Diamond pore network model used in this work. At each node, a pressure  $P_i$ is defined. In each link, 
the flow rate is a function of the pressure difference $\delta P=P_i-P_j$ according to a bi-viscous model.}
\end{figure}

We also choose a simple porous medium, a square lattice oriented at 45$^\circ$ with respect to the average flow direction, see
figure \ref{fig1}, consisting of $N_x$ links in the flow direction and $N_y$ links in the direction orthogonal to the flow direction.

\begin{figure}
\includegraphics[width=\hsize]{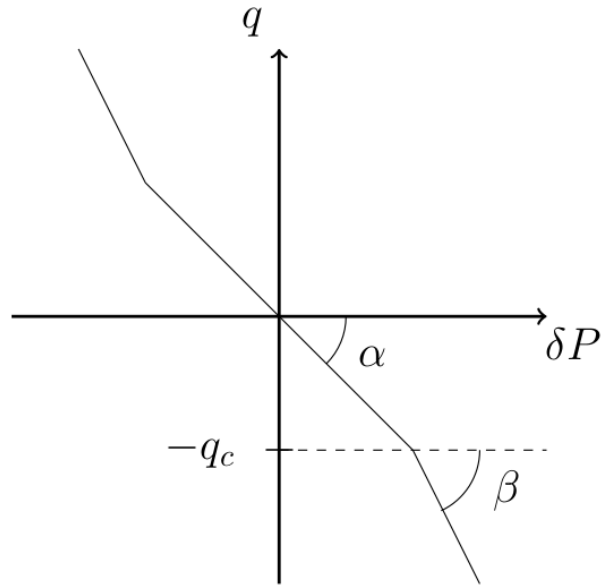}
\caption{\label{fig2} Bi-viscous flow curve. If the absolute value of the flow rate is below a local threshold $q_c$, the flow is linear with a mobility $\alpha$. Once the absolute value of the flow rate has reached the threshold the evolution is still linear but with different mobility 
$\beta$.}
\end{figure}

The constitutive equation for the fluid in a link in the lattice is given by   
\begin{equation}
\label{intro-1}
q(\nabla p) = \left\{
  \begin{array}{ll}
    -\alpha \nabla p & : |q| \le q_c\;,\\
    -\beta \nabla p +{\rm sgn}(q)\left[1-\frac{\beta}{\alpha}\right]q_c & : q_c \le |q| \;,\\
  \end{array}
\right.
\end{equation}
where  $q$ is the volumetric flow rate in the link, $\nabla p$ is the pressure drop across the link. There are three parameters,
$\alpha$, $\beta$ and $q_c$  The two first parameters, $\alpha$ and $\beta$ are mobilities when the fluid is either in the 
``$\alpha$-mode" or in the ``$\beta$-mode."  The third parameter, $q_c$ is the flow rate at which the fluid changes from
being $\alpha$-mode to $\beta$-mode.  We illustrate the constitutive equation in figure \ref{fig2}. To simplify the problem as 
much as possible, we let the two mobilities $\alpha$ and $\beta$ be the same for all links in the lattice.  However, each link
has its own flow rate threshold $q_c$ drawn from a probability distribution $p(q_c)$.

We will in the following study this system for arbitrary values of $\alpha$ and $\beta$ and for two threshold distributions; a uniform
distribution and an exponential distribution.  

In section \ref{symmetries}, we consider the symmetries inherent in the system.  There are two types of symmetries.  The first type is
related to what happens to the volumetric flow rate through the system, $Q$ when we scale the parameters.  Using the Euler theorem
for homogeneous functions we are able to write down the most general form of the volumetric flow rate. If we define $\langle q\rangle$
as $Q/N_y$, we find that $\langle q\rangle= \alpha\ \overline{q}(\nabla p,\beta/\alpha, \{q_c\}/\alpha)$, where $\{q_c\}$ refers
to the set of thresholds, one for each link.  The second type of symmetry is the self-duality of the square lattice leading to a mapping between the behavior of the system for a given ratio $\beta/\alpha$ and its inverse, $\alpha/\beta$.  Hence, we only need to discuss
$\beta/\alpha \ge 1$.   

We study in section \ref{fbm} the lattice with $N_x=1$, i.e., there is only one layer.  The model then becomes the {\it capillary
fiber bundle model\/} which is analytically tractable.  We find that for the uniform threshold distribution, the flow rate behaves as 
$\langle q\rangle-\langle q_c\rangle \sim (\nabla p-\nabla p_c)^2$ where $(\langle q_c\rangle,\nabla p_c)$ is a point only dependent
on the value of the ratio $\beta/\alpha$ and the limits of the uniform distribution $q_{\min}$ and $q_{\max}$.  This is reminiscent
of a {\it critical point.\/}  However, it is not a critical point.  There are no correlations developing in the system as 
$\nabla p$ approaches $\nabla p_c$.  Furthermore, the power law behavior is {\it not\/} seen when the threshold distribution is 
exponential.  

Section \ref{numerics} is devoted to the numerical algorithm we use to solve the flow patterns.  Our algorithm is based on the augmented Lagrangian algorithm, which we describe in this section.  

We present our results in section \ref{results}.  First we note that the two limits $\beta/\alpha\to 1$ and $\beta/\alpha\to \infty$, or
equivalently, $\beta/\alpha\to 0$ correspond to the {\it directed percolation\/} \cite{h00} and the {\it directed polymer\/} problems 
respectively \cite{hhr91}. This 
points us in the direction of there being a critical point in the problem in spite of the conclusion drawn for the capillary
fiber bundle model in section \ref{fbm}. Indeed, this is what we find: We find that $\langle q\rangle-\langle q_c\rangle \sim%
(\nabla p-\nabla p_c)^\mu$ where $\mu$ depends on the ratio $\beta/\alpha$ for the same type of treshold distribution that gave 
a power law dependence in the capillary fiber bundle model studied in section \ref{fbm}.  We define and measure a correlation length 
$L_{\max}\sim (\nabla p-\nabla p_c)^{-\nu}$.  The correlation length exponent $\nu$ also depends on the ratio $\beta/\alpha$.  In the 
limit $\beta/\alpha \to 1$, the longitudinal directed percolation correlation length exponent $\nu_\parallel=1.733847(6)$ \cite{j99}
is expected and our numerical results are consistent with this. In the directed polymer
limit $\beta/\alpha\to \infty$, however, the corresponding correlation length exponent is {\it not\/} the usual one, $\nu_\parallel=3/2$
\cite{rhh91}, but rather one that describes a {\it correlated directed percolation problem.\/}  

The last section \ref{summary} contains our summary and conclusions.

\section{Symmetries}
\label{symmetries}

In this section, we discuss the symmetries that lie hidden in the system we study, a square lattice of links obeying the
constitutive equation (\ref{intro-1}).  We consider two types of symmetry: one is based on scaling of the size and parameters
of the model.  Through the Euler theorem for homogeneous functions, we are able to write down the most general functional form
the volumetric flow rate through the network takes.  We then go on to exploring the geometrical symmetry inherent in the square
lattice due to self duality in the same way as first done by Straley \cite{s77}.  This symmetry demonstrates that we only need to explore 
the part of parameter space for which $\beta/\alpha\ge 1$.      

\subsection{Scaling symmetry}
\label{scaling}

The volumetric flow rate $Q$ shows a number of scaling symmetries.  We now combine these with the Euler theorem for homogeneous functions
to deduce the functional form of $Q=Q(\Delta P,\alpha,\beta,\{q_c\},N_x,N_y)$ \cite{hsbkgv18}.  Here $\{q_c\}$ is the set of thresholds, one for each link in the network.  The volumetric flow rate is extensive in the width of the network, $N_y$.  Hence,
\begin{eqnarray}
\label{scal-1}
&&Q(\Delta P,\alpha,\beta,\{q_c\},N_x,\lambda_y N_y)\nonumber\\
&=&\lambda_y Q(\Delta P,\alpha,\beta,\{q_c\},N_x,N_y)\;.
\end{eqnarray}
With respect to the length of the system, we find the symmetry
\begin{eqnarray}
\label{scal-2}
&&Q(\Delta P,\alpha,\beta,\{q_c\},N_x,N_y)\nonumber\\
&=&Q(\lambda_x\Delta P,\alpha,\beta,\{q_c\},\lambda_x N_x,N_y)\;.
\end{eqnarray}
A more subtle scaling symmetry is
\begin{eqnarray}
\label{scal-3}
&&Q(\Delta P,\lambda_q\alpha,\lambda_q\beta,\{\lambda_q q_c\},N_x,N_y)\nonumber\\
&=&\lambda_q Q(\Delta P,\alpha,\beta,\{q_c\},N_x,N_y)\;.
\end{eqnarray}
We also have the scaling symmetry
\begin{eqnarray}
\label{scal-4}
&&Q\left(\Delta P,\alpha,\beta,\{q_c\},N_x,N_y\right)\nonumber\\
&=&Q\left(\lambda_P\Delta P,\frac{\alpha}{\lambda_P},
\frac{\beta}{\lambda_P},\{q_c\},N_x,N_y\right)\;.
\end{eqnarray}
The length $N_x$ and width $N_y$ of the network are discrete variables.  By setting $\lambda_y=1/N_y$ we find from Equation 
(\ref{scal-1}) that 
\begin{eqnarray}
\label{scal-5}
&&Q(\Delta P,\alpha,\beta,\{q_c\},N_x,N_y)\nonumber\\
&=&N_yQ(\Delta P,\alpha,\beta,\{q_c\},N_x,1)\;.
\end{eqnarray}
The second scaling relation, Equation (\ref{scal-2}) gives when setting $\lambda_x=1/N_x$,
\begin{eqnarray}
\label{scal-6}
&&Q(\Delta P,\alpha,\beta,\{q_c\},N_x,N_y)\nonumber\\
&=&Q(\nabla p,\alpha,\beta,\{q_c\},1,N_y)\;,
\end{eqnarray}
where we have used the definition $\nabla p =\Delta P/N_x$. We now combine Equations (\ref{scal-5}) and (\ref{scal-6}) to get
\begin{eqnarray}
\label{scal-7}
&&Q(\Delta P,\alpha,\beta,\{q_c\},N_x,N_y)\nonumber\\
&=&N_yQ(\nabla p,\alpha,\beta,\{q_c\},1,1)=\langle q\rangle\;.
\end{eqnarray}
Hence, we define the average flow rate in the links as
\begin{equation}
\label{scal-8}
\langle q \rangle(\nabla p,\alpha,\beta,\{q_c\})=Q(\nabla p,\alpha,\beta,\{q_c\},1,1)\;.
\end{equation} 
This is thus an intensive variable with respect to the width and the length of the network.

\begin{figure}
\includegraphics[width=8.5truecm]{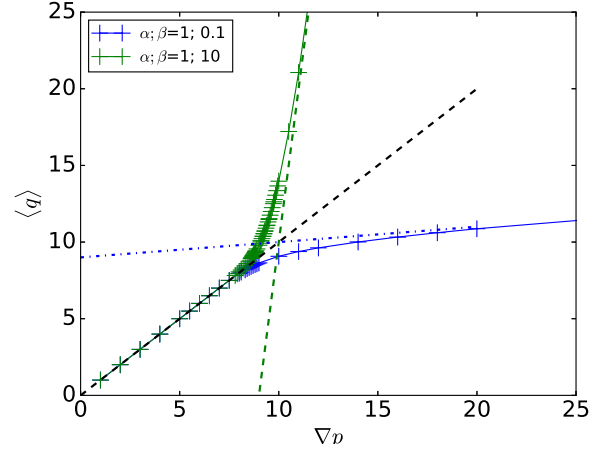}
\caption{\label{fig3} Example of the mean flow rate $\langle q \rangle$ as function of the mean gradient $\nabla p$ for two different bi-viscous model $(\alpha,\beta)=(1,10)$ (blue) and $(\alpha,\beta)=(1,0.1)$ (green).  As described in the text, the two cases are symmetrical through a dual mapping.}
\end{figure}

The two remaining scaling relations (\ref{scal-3}) and (\ref{scal-4}) involve continuous variables and we may thus make use of 
Euler's theorem for homogeneous functions. The Euler theorem is easy to implement for each of these four scaling symmetries: we take the derivative with respect to the scaling variable $\lambda$ in each expression and set the variable equal to one. 

The scaling relation (\ref{scal-3}) gives 
\begin{eqnarray}
\label{scal-9}
&&Q(\Delta P,\alpha,\beta,\{q_c\},N_x,N_y)\nonumber\\
&=&\left(\frac{\partial Q}{\partial \alpha}\right)\alpha
+\left(\frac{\partial Q}{\partial \beta}\right)\beta
+\sum_{\rm links} \left(\frac{\partial Q}{\partial q_c}\right) q_c\;,
\end{eqnarray}
or in terms of the intensive variable
\begin{eqnarray}
\label{scal-10}
&&\langle q\rangle(\nabla P,\alpha,\beta,\{q_c\})\nonumber\\
&=&\left(\frac{\partial \langle q\rangle}{\partial \alpha}\right)\alpha
+\left(\frac{\partial \langle q \rangle}{\partial \beta}\right)\beta
+\sum_{\rm links} \left(\frac{\partial \langle q\rangle}{\partial q_c}\right) q_c\;.
\end{eqnarray}
We define the functions
\begin{equation}
\label{scal-11}
A=-\left(\frac{\partial \langle q\rangle}{\partial \alpha}\right)\;,
\end{equation}
\begin{equation}
\label{scal-12}
B=-\left(\frac{\partial \langle q\rangle}{\partial \beta}\right)\;,
\end{equation}
and
\begin{equation}
\label{scal-13}
\{c\}=\left\{\left(\frac{\partial \langle q\rangle}{\partial q_c}\right)\right\}\;.
\end{equation}
There is one function $c$ for each link in the network. 

Whereas $\langle q\rangle$ is homogeneous of order one\footnote{A homogeneous function
$f(x,y)$ of order $n$ in the variables $x$ and $y$ fulfills the scaling relation $\lambda^n f(x,y)=f(\lambda x,\lambda y)$.} in the variables $\alpha$, $\beta$ and $\{ q_c\}$, the functions
$A$, $B$ and $\{c\}$ are homogeneous of order zero in these variables.  This means that the parameters $\alpha$, $\beta$ and $\{q_c\}$
only appear as ratios in these functions,   
\begin{equation}
\label{scal-14}
A=A\left(\nabla p,\frac{\beta}{\alpha},\frac{\{q_c\}}{\alpha}\right)\;,
\end{equation}
\begin{equation}
\label{scal-15}
B=B\left(\nabla p,\frac{\beta}{\alpha},\frac{\{q_c\}}{\alpha}\right)\;,
\end{equation}
and
\begin{equation}
\label{scal-16}
\{c\}=\left\{c\left(\nabla p,\frac{\beta}{\alpha},\frac{\{q_c\}}{\alpha}\right)\right\}\;.
\end{equation}
Equation (\ref{scal-9}) may thus be written
\begin{eqnarray}
\label{scal-17}
\langle q\rangle(\nabla P,\alpha,\beta,\{q_c\})&=&\nonumber\\
-A\left(\nabla p,\frac{\beta}{\alpha},\frac{\{q_c\}}{\alpha}\right)\alpha
&-&B\left(\nabla p,\frac{\beta}{\alpha},\frac{\{q_c\}}{\alpha}\right)\beta\nonumber\\
&+&\sum_{\rm links} c\left(\nabla p,\frac{\beta}{\alpha},\frac{\{q_c\}}{\alpha}\right) q_c\;.\nonumber\\
\end{eqnarray}

Scaling equation (\ref{scal-4}) combined with the Euler theorem gives
\begin{equation}
\label{scal-18}
\left(\frac{\partial Q}{\partial \Delta P}\right)\Delta P=\left(\frac{\partial Q}{\partial \alpha}\right)\alpha
+\left(\frac{\partial Q}{\partial \beta}\right)\beta\;,
\end{equation}
In terms of $\langle q\rangle$ and equation (\ref{scal-16}), we may rewrite this equation
\begin{eqnarray}
\label{scal-19}
&&m\left(\nabla p,\frac{\beta}{\alpha},\frac{\{q_c\}}{\alpha}\right)\nabla p\nonumber\\
&=&A\left(\nabla p,\frac{\beta}{\alpha},\frac{\{q_c\}}{\alpha}\right)\alpha\nonumber\\
&+&B\left(\nabla p,\frac{\beta}{\alpha},\frac{\{q_c\}}{\alpha}\right)\beta\;,
\end{eqnarray}
where we have defined the mobility
\begin{equation}
\label{scal-20}
m=-\left(\frac{\partial \langle q\rangle}{\partial \nabla p}\right)\;.
\end{equation}

From equations (\ref{scal-9}) and (\ref{scal-18}), we deduce that
\begin{equation}
\label{scal-21}
\langle q\rangle=\left(\frac{\partial \langle q\rangle}{\partial \nabla p}\right)\nabla p+\sum_{\rm links} c\ q_c
=-m\nabla p+\sum_{\rm links} c\ q_c\;,
\end{equation}
and with the help of equation (\ref{scal-19}) we find
\begin{eqnarray}
\label{scal-22}
\langle q\rangle=&-&a\left(\nabla p,\frac{\beta}{\alpha},\frac{\{q_c\}}{\alpha}\right)\alpha\nabla p\nonumber\\
&-&b\left(\nabla p,\frac{\beta}{\alpha},\frac{\{q_c\}}{\alpha}\right)\beta\nabla p\nonumber\\
&+&\sum_{\rm links} c\left(\nabla p,\frac{\beta}{\alpha},\frac{\{q_c\}}{\alpha}\right) q_c\;,
\end{eqnarray}
where we have defined
\begin{equation}
\label{scal-23}
a\left(\nabla p,\frac{\beta}{\alpha},\frac{\{q_c\}}{\alpha}\right)\nabla p
=A\left(\nabla p,\frac{\beta}{\alpha},\frac{\{q_c\}}{\alpha}\right)\;,
\end{equation}
and 
\begin{equation}
\label{scal-24}
b\left(\nabla p,\frac{\beta}{\alpha},\frac{\{q_c\}}{\alpha}\right)\nabla p
=B\left(\nabla p,\frac{\beta}{\alpha},\frac{\{q_c\}}{\alpha}\right)\;.
\end{equation}

We may take equation (\ref{scal-22}) one step further by dividing out the parameter $\alpha$, 
\begin{equation}
\label{scal-24-1}
\frac{\langle q\rangle}{\alpha} =\overline{q}\left(\nabla p,\frac{\beta}{\alpha},\frac{\{q_c\}}{\alpha}\right)\;,
\end{equation}
where
\begin{eqnarray}
\label{scal-24-2}
&&\overline{q}\left(\nabla p,\frac{\beta}{\alpha},\frac{\{q_c\}}{\alpha}\right)=\nonumber\\
&-&a\left(\nabla p,\frac{\beta}{\alpha},\frac{\{q_c\}}{\alpha}\right)\nabla p\nonumber\\
&-&b\left(\nabla p,\frac{\beta}{\alpha},\frac{\{q_c\}}{\alpha}\right)\frac{\beta}\alpha\ \nabla p\nonumber\\
&+&\sum_{\rm links} c\left(\nabla p,\frac{\beta}{\alpha},\frac{\{q_c\}}{\alpha}\right) \frac{q_c}{\alpha}\;.\nonumber\\
\end{eqnarray}
 
We may as a check, compare equation (\ref{scal-22}) --- our main result in this section --- with the constitutive equation (\ref{intro-1})
in the case when there is no disorder, i.e., when all $q_c$ are equal.  In this case, $\langle q\rangle$ should be equal to the
constitutive equation.  Hence, in this case we find,
\begin{equation}
\label{scal-25}
a\left(\nabla p,\frac{\beta}{\alpha}\right)=\Theta(q_c-|q|)\;,
\end{equation}
\begin{equation}
\label{scal-26}
b\left(\nabla p,\frac{\beta}{\alpha}\right)=\Theta(|q|-q_c)\;,
\end{equation}
and
\begin{equation}
\label{scal-27}
c\left(\nabla p,\frac{\beta}{\alpha}\right)=\Theta(|q|-q_c)\ {\rm sign}(q)\left(1-\frac{\beta}{\alpha}\right)\;.
\end{equation}
Here $\Theta$ is the Heaviside step function which is one for positive arguments and zero for negative arguments.  We note that if
$|q| < q_{\min}$, then equations (\ref{scal-25}) to (\ref{scal-27}) are correct as the disorder is not ``noticeable" in this flow regime. 

\subsection{Self-duality of the square lattice}
\label{duality}

\begin{figure}
\includegraphics[width=8.0truecm]{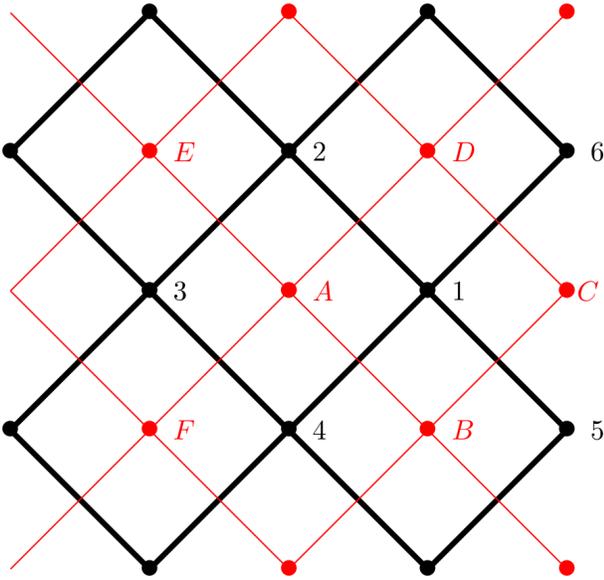}
\caption{\label{fig4} Sketch of the dual network construction. From the original network (black), one can construct a dual one (red), where the nodes are located at the center of the original cells. At each link of the dual network is associated a ``dual" flow rate
obtained from the pressure difference of the original network. At each node is associated a ``dual" pressure based on the original flow
rate. See the text for details.}
 \end{figure}

We define a dual network as sketched in Fig. \ref{fig4}.
A node is located at the center of each cell and there is  a link connecting each adjacent cell.
On each link, a "dual" current is defined from the pressure difference between pressure by the crossed link (from the original network),
\begin{eqnarray}
  j_{A \rightarrow B} &=& P_1 - P_4\;,\nonumber   \\
  j_{A \rightarrow D} &=& P_2 - P_1\;,\nonumber   \\
  j_{F \rightarrow A} &=& P_3 - P_4\;,\nonumber   \\
  j_{E \rightarrow A} &=& P_2 - P_3\;.
\label{d1}   
\end{eqnarray}
The current in the dual lattice satisfies the conservation of mass at each node (\emph{e.g.} Kirchhoff condition) since:
$ j_{A \rightarrow B} + j_{A \rightarrow D} - j_{F \rightarrow A} - j_{E \rightarrow A}=0 $

Moreover, one can define a pressure field $W$ on the dual lattice defined from this gradient,
\begin{eqnarray}
	W_A -  W_B &=& q_{1 \rightarrow 4 }\;,\nonumber \\
	W_B -  W_C &=& q_{1 \rightarrow 5 }\;,\nonumber \\
	W_C -  W_D &=& q_{1 \rightarrow 6 }\;,\nonumber \\
	W_D -  W_A &=& q_{1 \rightarrow 2 }\;. 
\end{eqnarray}
The definition is consistent once W is defined at a single point since the sum over a closed loop (and thus any) is equal to zero:
$(W_A -  W_B) + (W_B -  W_C) + (W_C -  W_D) + (W_D -  W_A)  = q_{1 \rightarrow 4 } + q_{1 \rightarrow 5 } +q_{1 \rightarrow 6 } + q_{1 \rightarrow 2 } =0.$

The ``dual" pressure gradient and current follows thus the constitutive equation:
\begin{equation}
	W_A - W_B = q_{1 \rightarrow 4 } =q( P_A - P_B  ) = q(  j_{A \rightarrow B} ).  
\end{equation}
Thus,
\begin{equation}
	j_{A \rightarrow B} = q^{-1}(W_A - W_B)
\end{equation}

The dual pressure and flow rate field satisfy thus the same kind of equation but with a local law which is  inverted.
It is important to note that the mean flow in the dual lattice is  perpendicular to the original one.

\section{Capillary fiber bundle model}
\label{fbm}

We now consider an analytically solvable model for the flow.  Let us assume that the network consists of a set of parallel links placed
between two fluid reservoirs kept at pressure $p=0$ and $p=\nabla p<0$, i.e., we are describing the {\it capillary fiber bundle model\/}
\cite{s53,s74,rhs19}. The constitutive equation for the fiber bundle is then given by
\begin{eqnarray}
\label{fbm-1}
Q&=&\sum_{i=1}^{N_y} \left[-\Theta(q_i-\alpha |\nabla p|) \alpha \nabla p\right.\nonumber\\
&-&\Theta(\alpha |\nabla p|-q_i)\beta\nabla p\nonumber\\
&+&\left.\Theta(\alpha|\nabla p|-q_i)\left(1-\frac{\beta}{\alpha}\right)q_i\right]\;,
\end{eqnarray}
where we have labeled the links according to their position, $i=1, N_y$ and $q_i$ is the threshold of the $i$th link.

Let us now relabel the links in ascending order with respect to their thresholds: $q_{(1)}\le q_{(2)} \le \cdots \le q_{(N_y)}$.  
Equation (\ref{fbm-1}) then becomes
\begin{eqnarray}
\label{fbm-2}
Q&=&\sum_{k=1}^{N_y} \left[-\Theta(q_{(k)}-\alpha |\nabla p|) \alpha \nabla p\right.\nonumber\\
&-&\Theta(\alpha |\nabla p|-q_{(k)})\beta\nabla p\nonumber\\
&+&\left.\Theta(\alpha|\nabla p|-q_{(k)})\left(1-\frac{\beta}{\alpha}\right)q_{(k)}\right]\;.
\end{eqnarray}

The thresholds are distributed according to the probability distribution $p(q_c)$, with a corresponding {\it cumulative probability\/} 
given by
\begin{equation}
\label{fbm-3}
P(q_c)=\int_0^{q_c} p(q)dq\;.
\end{equation}
According to order statistics, the mean value  of $k$th largest threshold --- mean value in the sense of averaging over an ensemble of
networks --- is given by
\begin{equation}
\label{fbm-4}
P(\overline{q}_{(k)})=\frac{k}{N_y+1}\approx \frac{k}{N_y}\;.
\end{equation}
Thus, the ensemble averages of the three types of sums in equation (\ref{fbm-2}) are then
\begin{equation}
\label{fbm-5}
\sum_{k=1}^{N_y}\Theta(\overline{q}_{(k)}-\alpha|\nabla p|)=N_y [1-P(\alpha|\nabla p|)]\;,
\end{equation}
\begin{equation}
\label{fbm-6}
\sum_{k=1}^{N_y}\Theta(\alpha |\nabla p|-\overline{q}_{(k)})=N_y P(\alpha|\nabla p|)\;,
\end{equation}
and
\begin{equation}
\label{fbm-7}
\sum_{k=1}^{N_y}\Theta(\alpha |\nabla p|-\overline{q}_{(k)})\overline{q}_{(k)}=N_y\ \int_0^{\alpha \nabla p} p(q)q\ dq\;.
\end{equation}
Inserted into equation (\ref{fbm-2}), these averages give
\begin{eqnarray}
\label{fbm-8}
\langle q\rangle &=&-[1-P(\alpha|\nabla p|)]\alpha\nabla p -P(\alpha|\nabla p|)\beta\nabla p\nonumber\\
&+&\left[1-\frac{\beta}{\alpha}\right]\int_0^{\alpha |\nabla p|} p(q)q\ dq\;,
\end{eqnarray}
where $\langle q\rangle=Q/N_y$.

\subsection{Uniform threshold distribution}
\label{fbm-uniform}

We now consider the concrete threshold distribution we will also employ in our numerical simulations on the square lattice: a uniform 
distribution on the interval $(q_{\min},q_{\max})$.  Hence,
\begin{equation}
\label{fbm-9}
p(q_c) = \left\{
  \begin{array}{ll}
    0 & : q_c \le q_{\min}\;,\\
    (q_{\max}-q_{\min})^{-1} & : q_{\min} < q_c < q_{\max}\;,\\
    0 & : q_{\max} \le q_c\;.\\
  \end{array}
\right.
\end{equation}
We define 
\begin{equation}
\label{fbm-10}
\nabla p_{\min}=-\frac{q_{\min}}{\alpha};,
\end{equation}
and
\begin{equation}
\label{fbm-11}
\nabla p_{\max}=-\frac{q_{\max}}{\alpha};.
\end{equation}
We also define
\begin{equation}
\label{fbm-12}
\nabla p_0=\frac{1}{2}\left[\nabla p_{\min}+\nabla p_{\max}\right]\;.
\end{equation}
Inserting these expressions into equation (\ref{fbm-8}) gives
\begin{equation}
\label{fbm-13}
\langle q\rangle =\left\{
   \begin{array}{ll}
  -\alpha \nabla p & : |\nabla p| \le |\nabla p_{\min}|\;,\\
   \frac{(\alpha-\beta)(\nabla p-\nabla p_c)^2}{2(\nabla p_{\max}-\nabla p_{\min})} &\\
  -\frac{\alpha(\alpha \nabla p_0-\beta\nabla p_{\min})}{\alpha-\beta} & : |\nabla p_{\min}| < |\nabla p|\\ 
                                                                       & : |\nabla p| < |\nabla p_{\max}|\;,\\ 
  -\beta\nabla p-(\alpha-\beta) \nabla p_0 & : |\nabla p_{\max}| \le |\nabla p|\;.\\
  \end{array}
\right.
\end{equation}
We have here defined
\begin{equation}
\label{fbm-14}
\nabla p_c=\frac{\alpha\nabla p_{\max}-\beta\nabla p_{\min}}{\alpha-\beta}\;.
\end{equation}
If we now define
\begin{equation}
\label{fbm-15}
\langle q_c\rangle=\frac{\alpha(\beta\nabla p_{\min}-\alpha \nabla p_0)}{\alpha-\beta}\;,
\end{equation}
we may cast the middle regime where $|\nabla p_{\min}| < |\nabla p| < |\nabla p_{\max}|$ in the form
\begin{equation}
\label{fbm-16}
\langle q\rangle=\langle q_c\rangle +\frac{(\alpha-\beta)}{2(\nabla p_{\min}-\nabla p_{\max})}\ (\nabla p-\nabla p_c)^2\;.
\end{equation}

It straight forward but somewhat tedious to rewrite the average flow rate $\langle q\rangle$, equation (\ref{fbm-13}) in the 
general form (\ref{scal-24-1}) and (\ref{scal-24-2}) resulting from the scaling relations (\ref{scal-1}) to (\ref{scal-4}).  
  
\subsection{Exponential threshold distribution}
\label{fbm-exponential}

Let us now consider the exponential threshold distribution
\begin{equation}
\label{fbm-17}
p(q_c)=\frac{e^{-q_c/q_0}}{q_0}\;,
\end{equation}
for $0 \le q_c <\infty$.  The corresponding cumulative distribution is
\begin{equation}
\label{fbm-18}
P(q_c)=1-e^{-q_c/q_0}\;.
\end{equation}
Inserted into equation (\ref{fbm-8}), this gives
\begin{eqnarray}
\label{fbm-19}
\langle q\rangle &=& -e^{\alpha\nabla p/q_0} \alpha \nabla p\nonumber\\
&-&\left(1-e^{\alpha\nabla p/q_0}\right)\beta\nabla p\nonumber\\
&+&\left[1-\frac{\beta}{\alpha}\right]\left[q_0-e^{\alpha\nabla p/q_0}\left(q_0-\alpha\nabla p\right)\right]\;,
\end{eqnarray}
where we are still assuming $\nabla p < 0$. Let us set $q_0=-\alpha \nabla p$. We then have the limits
\begin{equation}
\label{fbm-20}
\langle q\rangle = \left\{
   \begin{array}{ll}
  -\alpha \nabla p & : |\nabla p| \ll q_0/\alpha\;,\\
  -\beta\nabla p+(\alpha-\beta/) \nabla p_0 & : q_0/\alpha \ll |\nabla p|\;.\\
  \end{array}
\right.
\end{equation}
In contrast to the uniform distribution discussed in section \ref{fbm-uniform}, there is {\it not\/} a transitional regime between
the two limits of equation (\ref{fbm-20}) which is on the form (\ref{fbm-16}). 

Hence, the uniform distribution on an interval, (\ref{fbm-9}) results in $\langle q\rangle$ following a power law in $\langle q\rangle 
-\langle q_c\rangle$ vs.\ $\nabla p -\nabla p_c$, equation (\ref{fbm-16}), whereas the exponential distribution (\ref{fbm-17}) does not.
From the simple capillary fiber bundle model we may conclude that the power law behavior seen in equation (\ref{fbm-16}) is incidental 
and due to the uniform threshold distribution, which in itself is a power law.

We study a two-dimensional network mode in section \ref{results}. Surprisingly, we find that also in this case, only the uniform
distribution leads to a flow dependency on the pressure drop of the form
\begin{equation}
\label{fbm-21}
\langle q\rangle -\langle q_c\rangle \sim (\nabla p -\nabla p_c)^\mu\;.
\end{equation}
In this case, however, the exponent $\mu$ depends on the parameter ratio $\beta/\alpha$.   

\section{Numerical method: Augmented Lagrangian}
\label{numerics}

For completeness, this section describes the numerical method used to solve the non-linear  Kirchhoff equation.
This section is not required to understand the results that follow. 

The method used is based on the Augmented Lagrangian method commonly used to solve the Stokes equation for yield stress fluids 
\cite{dl76,gl89}. It is based on a variational method of the problem.
Indeed, if we rewrite the local equation \eqref{intro-1} and introduce the function $f(q)$ as :
\begin{equation}
\delta p(q) = -f(q)=\left\{
\begin{array}{ll}
  - \frac{1}{\alpha} q & :  |q|< q_c\; \\ 
  - \frac{1}{\beta}  \left[ q - \frac{q}{|q|} ( 1 - \frac{\beta}{\alpha} ) \right] &  :  |q|>q_c\;,\\ 
\end{array}
\right.
\end{equation}
We define the function $\phi(q) =\int_0 ^q f(q') dq'$.
The flow field $\{ q_l\}$ solution of equation \eqref{intro-1}, with the constraints of imposed inlet and outlet pressures at the boundaries $p_{in}$ and $p_{out}$, can be written as the saddle point of the functional
\begin{eqnarray}
&&\max_{ \{ \lambda_n \} } \min_{\{q_l\}} \Phi[\{ q_l \}, \{ \lambda_n\} ]\nonumber\\ 
&=&\sum_{l\in \mathcal{L} } \left[ \phi(q_n) - \delta_{l,in} p_{in} q_l   + \delta_{l,out} p_{out} q_l  \right]\nonumber\\ 
&+&\sum_{n \in \mathcal{N}} \lambda_n \sum_{l'\in {\mathcal{V} }(n)} q_{l'},
 \end{eqnarray}
 where $\mathcal{L}$ represents the ensemble of links,  $\mathcal{N}$ the ensemble of nodes and $\mathcal{V}(n)$ the ensemble of links connected to the node $n$.
The symbol $\delta_{l,in}$ (resp.\ $\delta_{l,out}$) is equal to $1$ if the link is connected to the inlet (resp.\ outlet) node and 
to $0$ otherwise. The $\{\lambda_n\}$ field is a set of Lagrangians which impose the conservation of mass at each node (and 
it can thus be associated to a pressure field).

The main idea of the Augmented Lagrangian method is to introduce a secondary set of velocities $\{j_l\}$ to decouple the nonlinear rheology to the Kirchhoff equation. 
Another constrain is then added $\{j_l\} = \{q_l\}$ via the Lagrangian method.

The velocity field is thus the solution of
\begin{eqnarray}
\label{eq:AL_functional}
&&\max_{ \{ \lambda_n \},\{ \mu_n\} } \min_{\{q_l\},\{ j_l \}} \Psi[\{ q_l \}, \{ j_l \}, \{ \lambda_n\}, \{ \mu_l\} ]\nonumber\\
&=& \sum_{l\in \mathcal{L} } \left[ \phi(q_n)
-\delta_{l,in} p_{in} j_l + \delta_{l,out} p_{out} j_l\right.\nonumber\\
&+&\left.\mu ( j_l - q_l) +  \frac{\epsilon}{2} | q_l - j_l|^2
 \right]\nonumber\\ 
&+& \sum_{n \in \mathcal{N} } \lambda_n \sum_{l'\in \mathcal{V}(n)} j_{l'}\;,
\end{eqnarray}
where $\{ \mu_l \}$ is a Lagrangian set. The quadratic term is an additional penalty term which characterizes the augmented Lagrangian  approach, where $\epsilon$ is a parameter.

The methods consists now in implementing an iterative algorithm to reach the saddle point starting from an initial guess $\{q^0_l\}$, $\{ j^0_l \}$, $\{ \lambda^0_n \}$ and $\{ \mu^0_l\}$.

Knowing $\{q^{n}_l\}$, $\{ j^{n}_l \}$, $\{ \lambda^{n}_n \}$ and $\{ \mu^{n}_n\}$, the algorithm is decomposed in the following steps:

\paragraph*{Determination of $\{ j^{n+1}_l \}$ and $\{\lambda^{n+1}_n \}$:}
For this, we should solve:
\begin{eqnarray}
& & \forall l\in \mathcal{L}, \frac{\partial}{\partial j_l} \Psi[\{ q^{n}_l \}, \{ j_l \}, \{ \lambda_n\}, \{ \mu^{n}_l\} ] =0\;,
\nonumber \\
& & \forall n\in \mathcal{N}, \frac{\partial}{\partial \lambda_n} \Psi[\{ q^{n}_l \}, \{ j_l \}, \{ \lambda_n\}, \{ \mu^{n}_l\} ] =0\;,
\end{eqnarray}
which reads
\begin{eqnarray}
&& \forall l\in \mathcal{L}, j^{n+1} = - \frac{1}{\epsilon} ( \lambda_{l+}^{n+1} - \lambda_{l-}^{n+1} + \mu_l^{n} - \epsilon q_l^{n} ) \\
&& \forall n\in \mathcal{N},  \sum_{l'\in \mathcal{V}(n)} j_{l'}^{n+1} = 0, 
\end{eqnarray}
where $\lambda^{n+1}_{l+}$ and  $\lambda^{n+1}_{l-}$ are the lagrangian of the two nodes adjacent  to the link $l$. For nodes adjacent to the outlet (resp. inlet), $\lambda_+$ (resp. $\lambda_-$) has to be replaced with $p_{out}$ (resp. $p_{in}$).

The most important point of this set of equations is that it is equivalent to solving the standard linear Kirchhoff equation with a constant permeability $1/\epsilon$ but with an additional source term $\mu_l^{n} - \epsilon q_l^{n}$. It can thus be solved by standard linear methods (Cholesky, LU decomposition, etc.).

\paragraph*{Determination of $q_l^{n+1}$:} 
We  solve
\begin{equation}
 \forall l\in \mathcal{L}, \frac{\partial}{\partial q_l} \Psi[\{ q_l \}, \{ j^{n+1}_l \}, \{ \lambda^{n+1}_n\}, \{ \mu^{n}_l\} ] =0,
\end{equation}
yielding to the local, but implicit equation:
\begin{equation}
\forall l\in \mathcal{L}, \phi'( q^{n+1}_l) + \epsilon q^{n+1}_l = \mu + \epsilon j^{n+1}_l
\end{equation}
Noting $y=\mu + \epsilon j^{n+1}_l$, the solution is:
\begin{equation}
	q^{n+1}_l = \left\{
	\begin{array}{ll}
	\frac{1}{1/\alpha + \epsilon} y & : |y|<\epsilon i_c + \frac{i_c}{\alpha}\\
	\frac{1}{1/\beta + \epsilon} &\\  
        \left[ |y| + (1/\beta-1/\alpha)i_c \right] \rm{sign}(y) & : |y|>\epsilon i_c + \frac{i_c}{\alpha}\\
	\end{array}
	\right.
\end{equation}
\paragraph*{Determination of $\mu_l^{n+1}:$} 
For this we update in the direction of the gradient (Newton method):
\begin{equation}
	\mu_l^{n+1} = \mu_l^{n+1} + \gamma ( j^{n+1}_l - q^{n+1}_l ), 
\end{equation}
where $\gamma$ is a parameter set to $\gamma=\epsilon$ for simplicity.

In practice, this algorithm iterated until the relative variation of the total flow rate between two step is below $10^{-5}\%$. The computational time and the number of steps is strongly varying depending on $\beta$ but also on the applied pressure.

\section{Results}
\label{results}

We now our numerical model based on the network show in figure \ref{fig1} and the algorithm described in section \ref{numerics}. We use the
link threshold distribution (\ref{fbm-9}) with $q_{\min}=7.5$ and $q_{\max}=12.5$ in the following.   

\subsection{Criticality}
\label{criticality}

As noted above, due to the distribution of thresholds, the links will reach their threshold at different macroscopic pressures. A link $l$ will be defined as being in $\beta$-mode if $q_{l}>q_c$ and in $\alpha$-mode otherwise.
Similar to the percolation problem, a macroscopic change in flow regime is expected once the existence of percolation pathways of 
$\beta$-mode links. However, it is important to note a major difference with the percolation problem which lies in the fact that the mode
of a link influences the neighboring links.
Indeed, in the case of $\beta>\alpha$, once a link switches to $\beta$-mode, the flow will be easier through it. This will
tend to concentrate the flow towards it. It will therefore increase the flow in the  upstream and downstream neighboring links and  therefore pushes these links towards the $\beta$-mode.
In the opposite case, for $\beta<\alpha$, the $\beta$-mode have a lower conductivity once entering this mode compared to what it would have in $\alpha$-mode. Flow will therefore tend to go around it, increasing the flow in the lateral links. Consequently, $\beta$-mode links 
will tend to correlate in the streamwise (or lateral) direction for 
$\beta>\alpha$ and orthogonally to the streamwise direction for  $\beta<\alpha$) \cite{wbhh96}.

The intermediate case $\beta=\alpha$ is interesting as the mode of a link has no influence on its neighbors.
Since the mobility are the same for every link, the flow rate and the pressure gradient become homogeneous and equal to the mean flow rate and mean gradient.  The problem is therefore identical to the {\it directed percolation problem\/} \cite{h00}.

The other limit $\beta/\alpha \gg 1$, the problem becomes identical to a yield stress fluid \cite{hhr91,grhbtc90,tb13}. 
The critical path is then related to the directed polymer problem \cite{hhr91,kz87,taph13}, as it corresponds to the path that minimize the sum of local pressure threshold $\Delta P_c = \min \sum (q_c/\alpha)$.   

\subsection{Pathscape method} 
\label{pathscape}

\begin{figure}
 \includegraphics[width=7.5truecm]{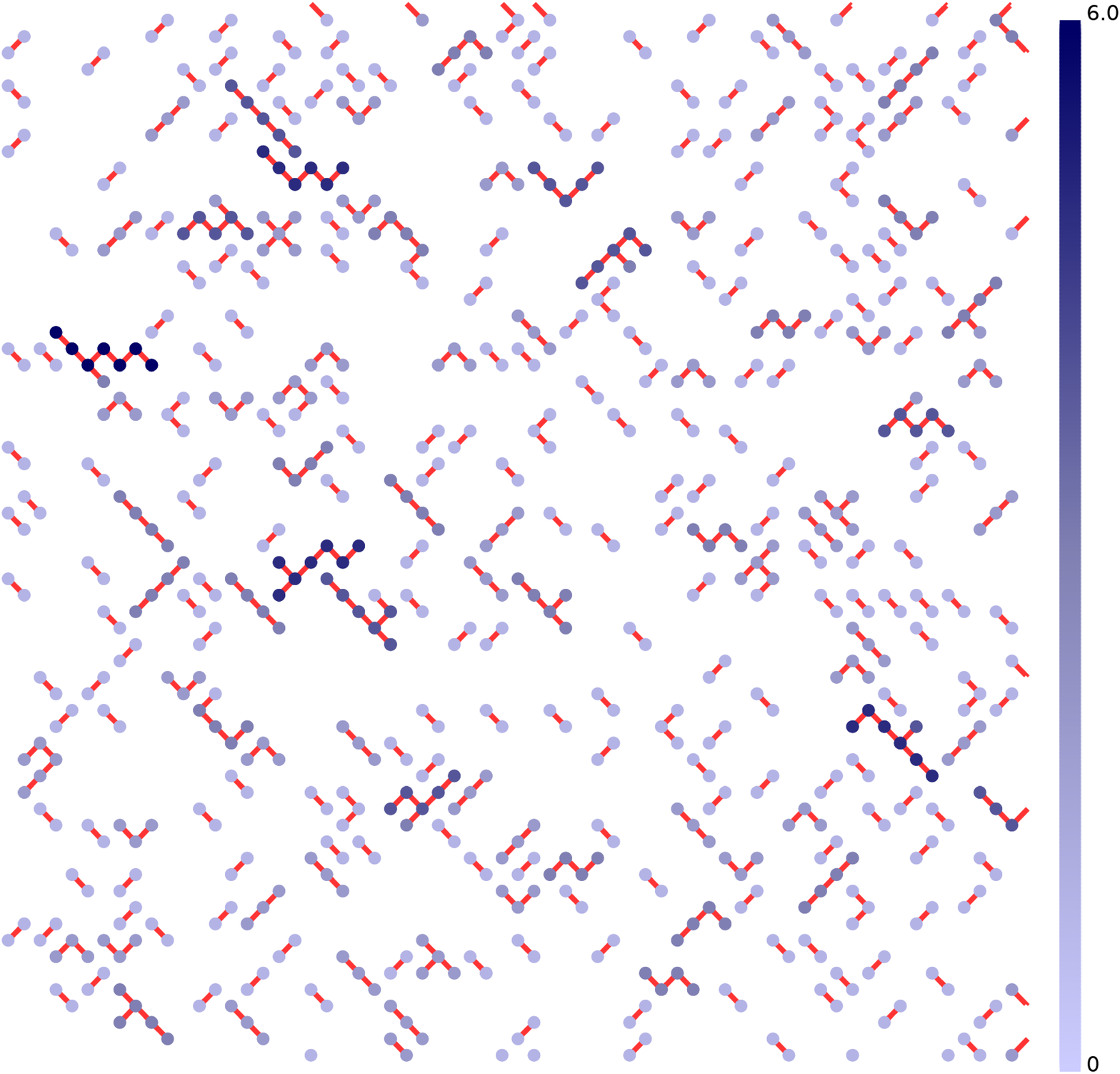}
 \includegraphics[width=7.5truecm]{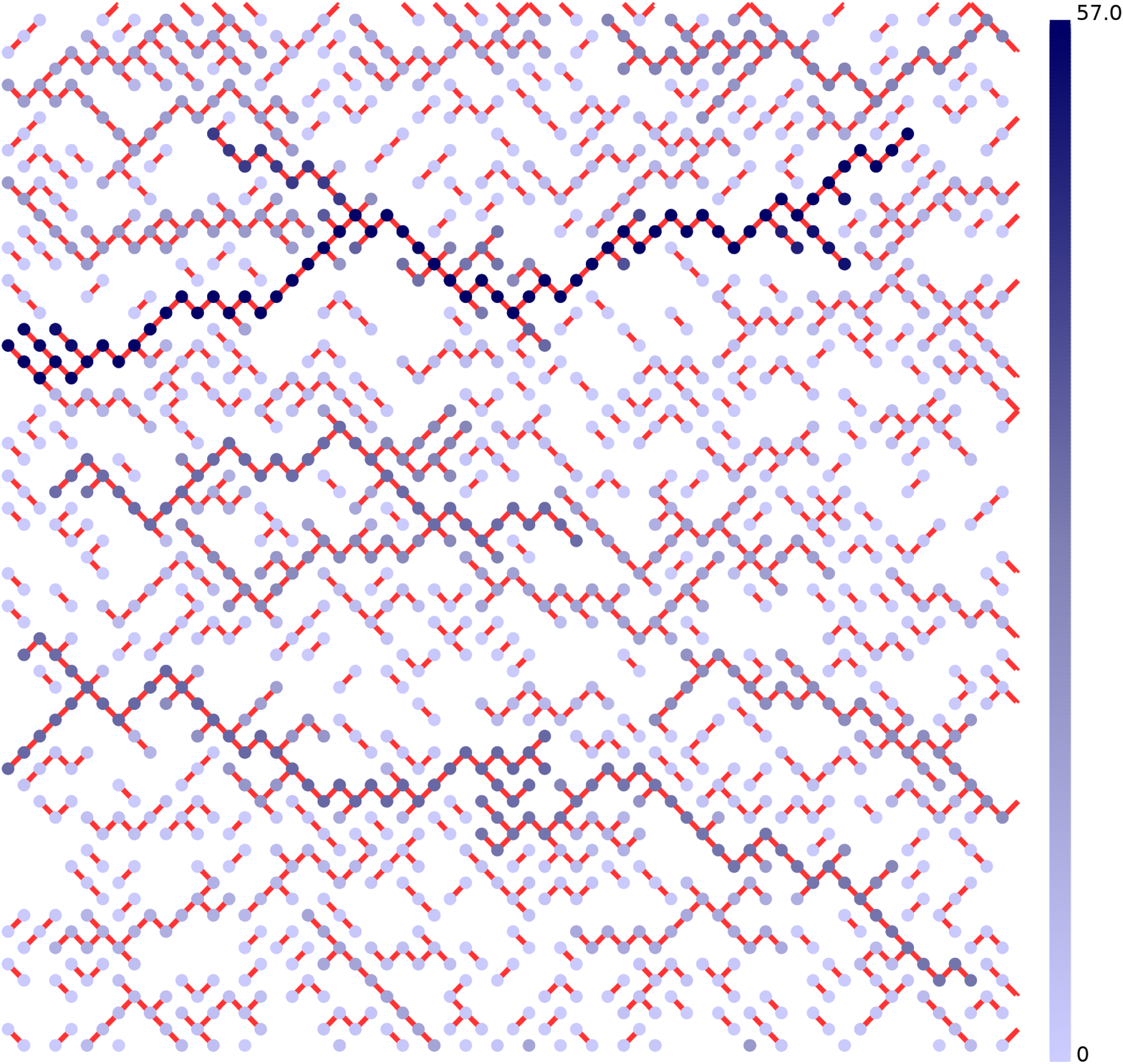}
\caption{\label{fig5} Pathscapes in the network at pressure differences $\nabla p=8$ (upper) and $\nabla p=8.6$ (lower).
The links in $\alpha$-mode are not shown.  Each link in $\beta$-mode have been assigned a color. The color reflects the
length of the path the link in $\beta$-mode belongs to according to the bar to the right of each network. 
The shortest paths are light blue, the longest are dark blue.}
\end{figure}

To quantify this phenomenon and to determine the percolation pressure, we want to determine the longest directed path of the $\beta$-mode links.  This quantity is essentially the longitudinal correlation length in directed percolation \cite{j99}.  We map the length of all
paths by invoking a pathscape approach as described in \cite{taph13} for yield-stress fluids.

We introduce the node field $L_{n}$  representing the longest upstream directed path ending at $n$.
$L_n$ can be determined from a transfer matrix algorithm propagating from left to right (stream direction).
If we note, at a given node $n$,  $l_1$ and $l_2$ the two upstream neighbor links and $n_1$ and $n_2$ the corresponding nodes.  We associate
binary variables $m_1$ and $m_2$ with the two links $l_1$ and $l_2$.  If link $l_1$ is in $\beta$-mode, then $m_1=1$, otherwise $m_1=0$ ---
and likewise for link $l_2$.  We then have that 
\begin{equation}
\label{linktrans}
L_n=\max\left[(L_{n_1}+1)m_1,(L_{n_2}+1)m_2\right]\;.
\end{equation}
We then construct  the node field $R_{n}$ containing the longest directed path ending at $n$ but propagating in the downstream direction. The algorithm is identical to the previous one but it propagates in the upstream direction from the rightmost column.

\begin{figure}
\includegraphics[width=8.5truecm]{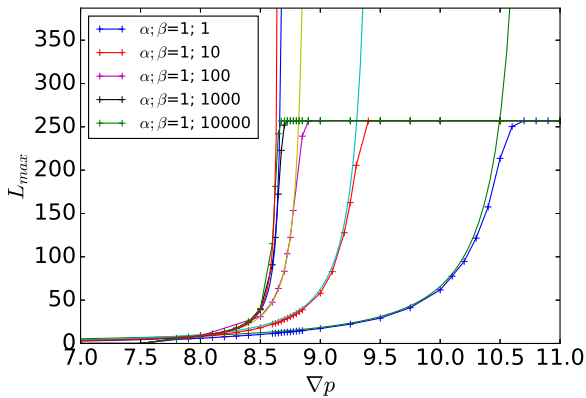}
\includegraphics[width=8.5truecm]{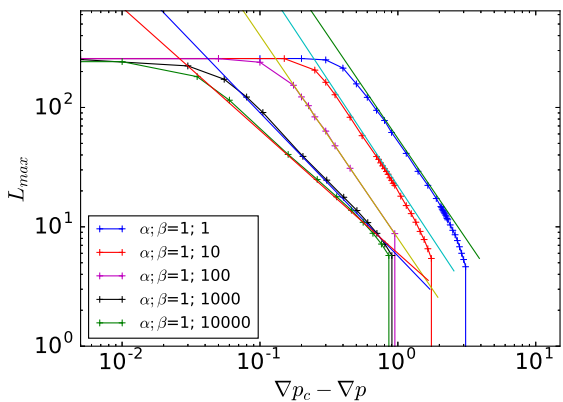}
\caption{\label{fig6} Correlation length $L_{\max}$ as function of  the gradient of pressure $\nabla p$ (a) or of the distance to the critical pressure $|\nabla p-\nabla p_c|$ (b) for different value of $\beta$.
The solid line correspond to the power law fit given by equation \eqref{eq:Lmax}.
The system size is $256\times256$.}
\end{figure}

Once both fields have been determined, we sum the two to obtain the pathscape $T_n=L_{n}+R_{n}$, which contains the length of longest directed percolating path passing by the node $n$. 
From this pathscape, we can then identify the longest directed path $L_{max}=\max(T_n)$
In figure \ref{fig5}, we present two examples of such a pathscape at two different imposed pressure. 
In this figure, we can see the longest cluster path in dark blue. At low applied pressure, the longest cluster is quite low $L_{\max}=7$, whereas at higher pressure, $L_{\max}$ is closer to the system size.

It is important to note that the pathscape we have defined here is {\it not\/} the landscape of minimal paths \cite{taph13}. In the limit
$\beta\to\alpha$ the pathscape reflects the clusters in directed percolation as noted in section \ref{criticality}. However, when 
$\beta \neq\alpha$, the paths we identify correspond to directed percolation clusters.  However, the directed percolation is now 
{\it correlated.\/}  

\subsection{Evolution of the correlation length $L_{\max}$}
\label{evolcorr}

\begin{figure}
 \includegraphics[width=8.5truecm]{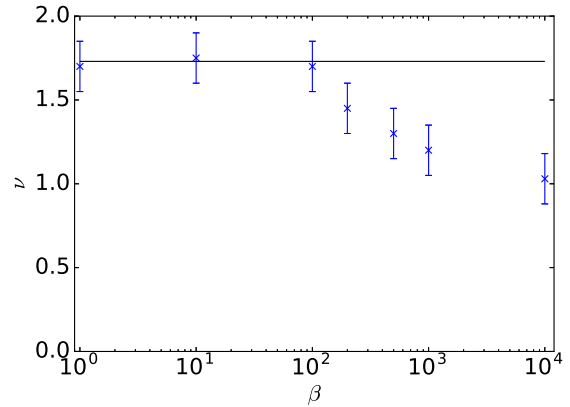}
 \includegraphics[width=8.5truecm]{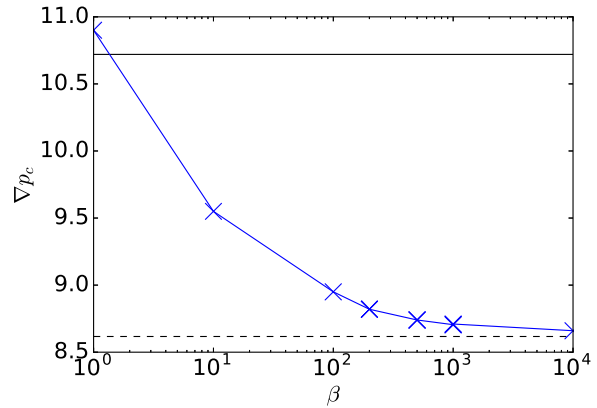}
\caption{\label{fig7}Upper panel:$\nu$ as function of $\beta$. The horizontal line corresponds to the directed percolation exponent 
$\nu\approx 1.72$.
Lower panel: Critical gradient of pressure $\nabla P_c(\beta)$ as function of $\beta$. The upper line corresponds to directed percolation 
($p_c=0.644700185(5)$ \cite{j99}). The line below (dashed) corresponds to the average of the directed polymer algorithm. 
The system size is $256 \times 256$}
\end{figure}

In figure \ref{fig6}, we investigate the evolution of $L_{\max}$ as function of the applied pressure.
As it can be seen, the correlation length increases with pressure until it reaches the system length $N_x$.
Similarly to percolation, one can see in figure \ref{fig6}(b) that the correlation length diverges as a power law close to a critical
pressure gradient $\nabla p_c$,
\begin{equation}
\label{eq:Lmax}
L_{\max} \propto ( \nabla p_c -  \nabla p )^{-\nu}.
\end{equation}
We note  in this figure that the exponent $\nu$ seems to vary with $\beta$.
In figure \ref{fig7}, we display the evolution of $\nu$ and the critical pressure gradient $\nabla p_c$ against the parameter $\beta$.
As we can see, $\nu$ and $\nabla p_c$ decrease significantly with $\beta$.
Where the limit $\beta \rightarrow 1$ is consistent with the results found in the literature on directed percolation, $\nu = \nu_\parallel =1.733847(6)$ \cite{j99}.  Our best estimate of the threshold pressure is $\nabla p_c \approx 10.72$.

At the end of section \ref{pathscape} we noted that the pathscape we have identified is {\it not\/} related to the pathscape spanned
by minimal paths in the limit $\beta/\alpha\to\infty$.  If that were the case, we would have expect $\nu$ to approach the value 
$\nu_{\parallel}=3/2$ \cite{rhh91}.  Rather, we are identifying directed percolation clusters in a correlated landscape, and this directed
percolation $\nu$ is approaching the value 1 in this limit.   

\subsection{Flow curve}
\label{flow}

\begin{figure}
 \includegraphics[width=8.5truecm]{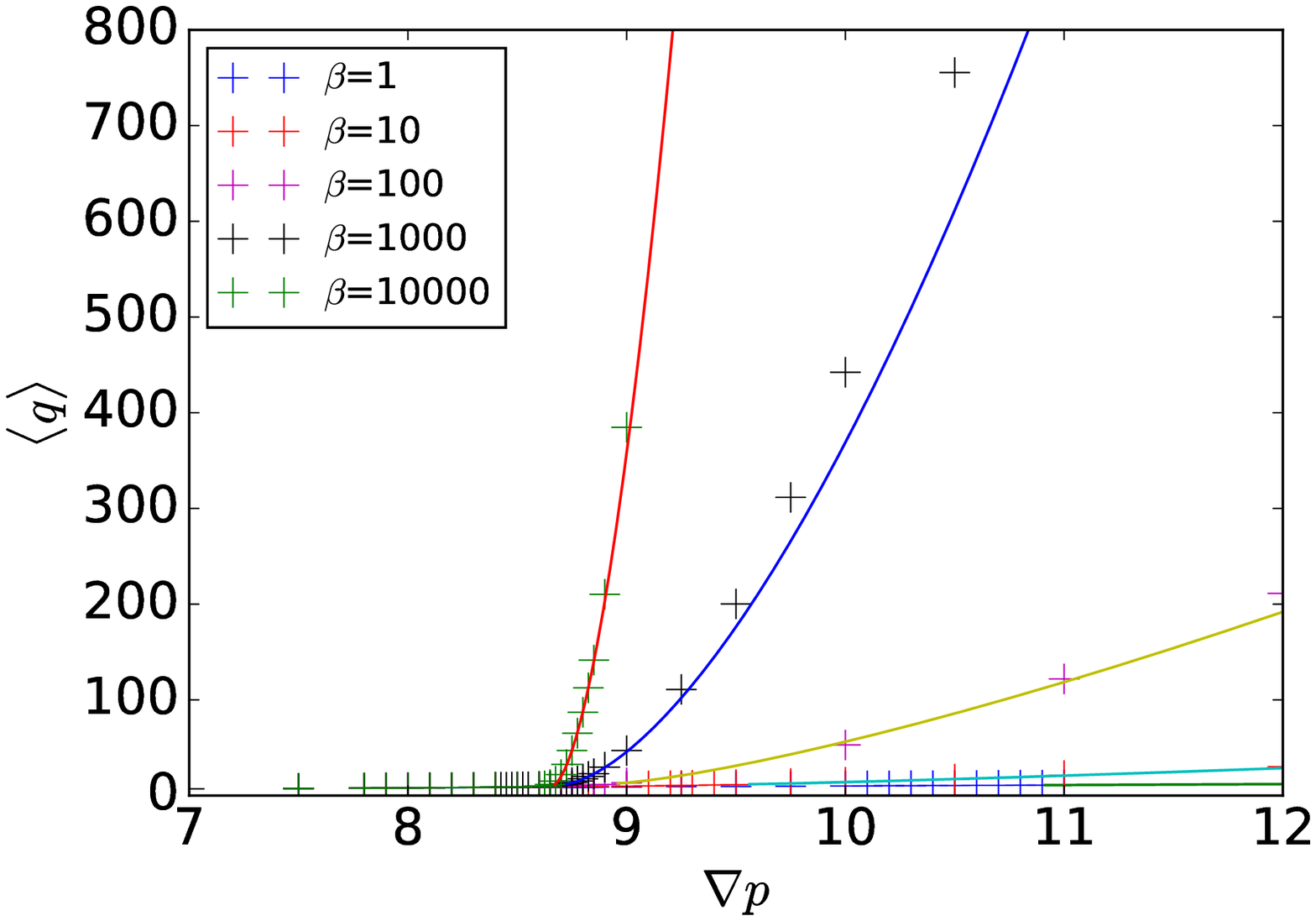}
 \includegraphics[width=8.5truecm]{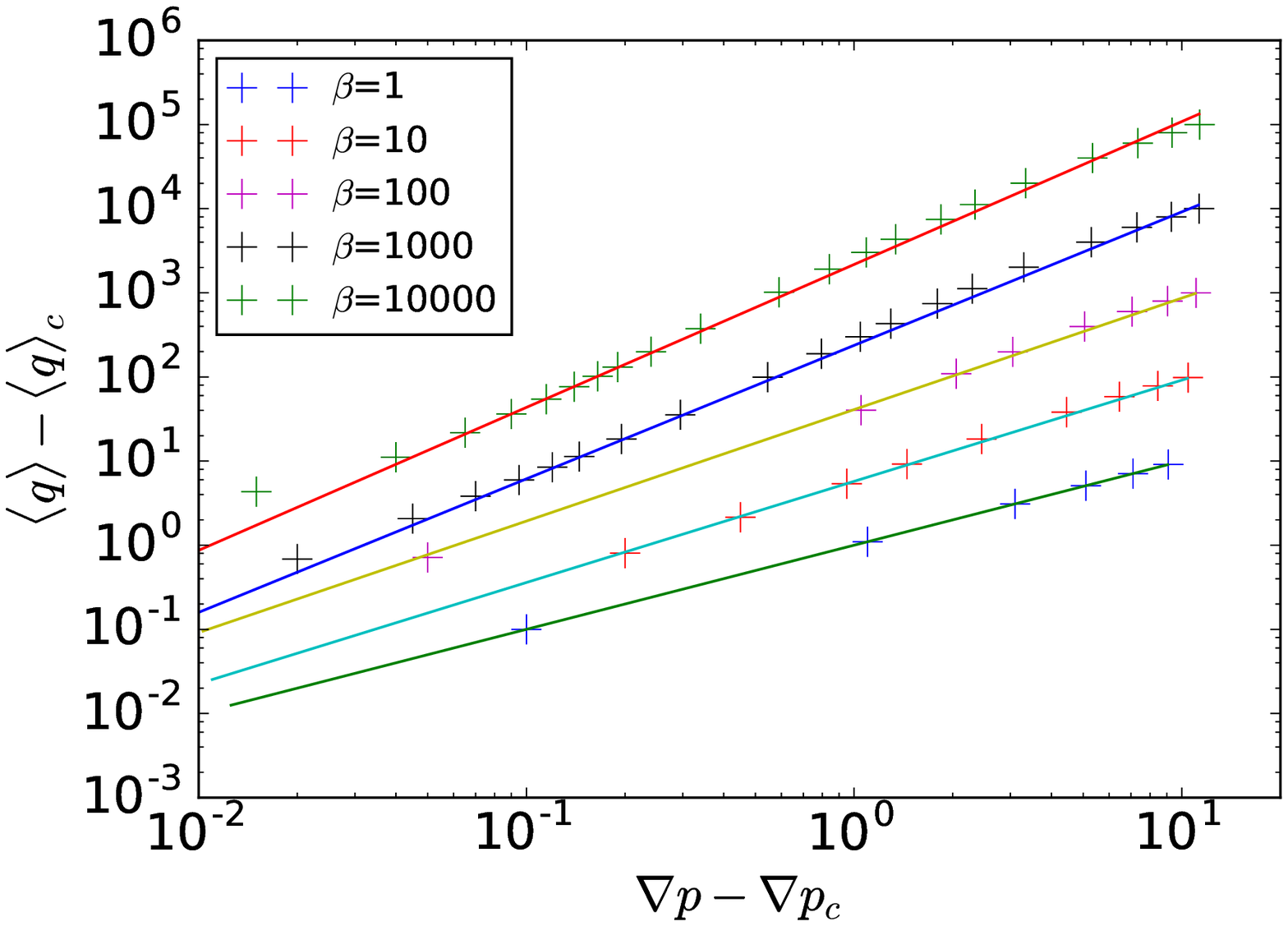}
\caption{\label{fig8} Mean flow rate $\langle q \rangle$ as function of the mean pressure gradient (upper panel)) 
and of the distance to the critical pressure gradient $\nabla p - \nabla p_c$ (lower panel) for different $\beta$.
The solid lines correspond the power law fit given by equation \eqref{eq:Q_P}. The system size is $256 \times 256$.}
\end{figure}

\begin{figure}
\begin{center}
 \includegraphics[width=0.6\textwidth]{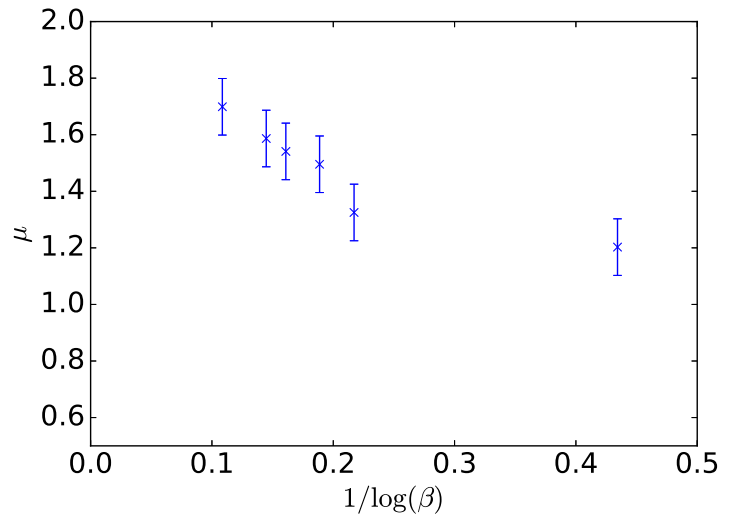}
 \end{center}
	\caption{ \label{fig9} Flow exponent $\mu$ as function of $1/\log(\beta)$ for a system size $256 \times 256$.}
\end{figure}

We now investigate the flow curve. Figure \ref{fig8} displays the evolution  of the  mean flow rate as function of the pressure gradient and for different $\beta$. 
In the lower figure, we show that, close to the critical pressure, the flow rate follows also a power-law which can be written in the form:
\begin{equation}
\label{eq:Q_P}
	\langle q \rangle - \langle q \rangle_c \propto (\nabla p - \nabla p_c)^\mu\;,
\end{equation}
where $q_c$ is a constant obtained by interpolating the data at the critical pressure.
Here also  we remark that the exponent $\mu$ varies with the the coefficient $\beta$.
In figure \ref{fig9}, we report the evolution of this exponent as a function of $1/\log(\beta)$.  For $\beta=\alpha=1$ we have the
obvious limiting value $\mu=1$.  As $\beta$ increases, so does the value of $\mu$.  By plotting $\mu$ against $1/\log(\beta)$ we may 
estimate the limiting value for $\beta\to\infty$, which is not inconsistent with the value $\mu=2$; the value suggested by Roux
and Herrmann in 1987 \cite{rh87}.  

We note that the functional form $\langle q\rangle$, equation (\ref{eq:Q_P}), based on the uniform threshold distribution (\ref{fbm-9}),
gives a behavior closely related to the one found for the capillary fiber bundle model with the same type of threshold
distribution, see equation (\ref{fbm-16}), but with $\mu=2$.  The correlation length exponent $\nu$ cannot be defined in the
capillary fiber bundle model. 

In section \ref{fbm-exponential}, we studied the capillary fiber bundle model with an exponential threshold distribution 
(\ref{fbm-17}).  We have used the same distribution for the network model considered here.  As in the capillary fiber bundle 
model, we do not find a power law of the type (\ref{eq:Q_P}) in this case, nor do we find a power law for the correlation
length, (\ref{eq:Lmax}).    

\section{Summary and Conclusions}
\label{summary}

We have in this paper explored the behavior of a bi-viscous fluid moving in a square lattice subject to the constitutive
equation (\ref{intro-1}) for each link. This system contains a critical point which leads to the behavior 
$\langle q\rangle-\langle q_c\rangle \sim (\nabla p-\nabla p_c)^\mu$ for the volumetric flow rate and 
$L_{\max}\sim (\nabla p-\nabla p_c)^{-\nu}$ for the correlation length when a uniform threshold distribution is used.  
However,
the two limits of the ratio between the two parameters representing the mobilities, $\beta/\alpha\to 1$ and $\beta/\alpha\to \infty$, or equivalently, $\beta/\alpha\to 0$ correspond to the {\it percolation\/} and the {\it directed polymer\/} problems respectively. These 
{\it are\/} problems containing critical points.

There are still a number of open questions concerning this system. We list them as follows:
\begin{itemize}
  \item We have only considered $\nabla p \ge \nabla p_c$.  What happens on the other side of the critical point?
  \item The critical exponents $\mu$ and $\nu$ are functions of the parameter ratio $\beta/\alpha$.  Is this a crossover
        or are we dealing with a non-universal exponent?  
  \item We have only dealt with $\beta \ge 0$.  What happens for $\beta < 0$?  The limit $\beta\to -\infty$ turns the model
        into the fuse model.  What happens when $\beta$ is barely negative?  Our numerical algorithms is not capable of
        handling this problem.
  \item It would be more realistic, but also more challenging to consider a power-law type characteristic for the constitutive 
        equation for $q\ge q_c$.  How will this change our conclusions?
  \item Why do we not see critical behavior for the exponential threshold distribution in the network model?
\end{itemize}


We thank the Research Council of Norway through its Centres of Excellence funding scheme, project number 262644. AH thanks the 
Universit{\'e} de Paris-Sud for funding through a visiting professorship.



\begin{thebibliography}{10}

\bibitem{c72} P.\ J.\ Carreau, Rheological equations from molecular network theories, Trans.\ Soc.\ 
Rheol.\ {\bf 16}, 99 (1972), doi.org/10.1122/1.549276.

\bibitem{hb26} W.\ H.\ Herschel and R.\ Bulkley, Konsistenzmessungen von Gummi-Benzoll{\"o}sungen, Kolloid Zeitschrift, {\bf 39} 291-–300
(1926), doi.org/10.1007/BF01432034.

\bibitem{w96} S.\ Whitaker, The Forchheimer equation: A theoretical development, Transp.\ Porous Med.\ {\bf 25}, 27-–61
(1996), doi.org/10.1007/BF00141261.

\bibitem{sh12} S.\ Sinha and A.\ Hansen, Effective rheology of immiscible two-phase flow in porous media, EPL {\bf 99}, 44004
(2012), doi.org/10.1209/0295-5075/99/44004.

\bibitem{wb36} R.\ D.\ Wyckoff and H.\ G.\ Botset, The flow of gas-liquid mixtures through unconsolidated sands, J.\ Appl.\ Physics,
{\bf 7}, 325 (1936), doi.org/10.1063/1.1745402.

\bibitem{rhg87} S.\ Roux, A.\ Hansen and E.\ Guyon, Criticality in non-linear transport properties of heterogeneous materials,
J.\ Phys.\ France {\bf 48}, 2125--2130 (1987), doi.org/10.1051/jphys:0198700480120212500. 

\bibitem{hrh90} E.\ L.\ Hinrichsen, S.\ Roux and A.\ Hansen, The conductor-superconductor transition in disordered superconducting materials, Physica C, {\bf 167}, 433--455 (1990), doi.org/10.1016/0921-4534(90)90364-K.

\bibitem{s77} J.\ P.\ Straley, Critical exponents for the conductivity of random resistor lattices, Phys.\ Rev.\ B {\bf 15}, 5733 (1977),
doi.org/10.1103/PhysRevB.15.5733.

\bibitem{h00} H.\ Hinrichsen, Non-equilibrium critical phenomena and phase transitions into absorbing states, Adv.\ Phys.\ 
{\bf 49}, 815 (2000), doi.org/10.1080/00018730050198152.

\bibitem{hhr91} A.\ Hansen, E.\ L.\ Hinrichsen and S.\ Roux, Roughness of crack interfaces, Phys.\ Rev.\ Lett.\ {\bf 66}, 2476 (1991),
doi.org/0.1103/PhysRevLett.66.2476.

\bibitem{j99} I.\ Jensen, Low-density series expansions for directed percolation: I. A new efficient algorithm with applications to the square lattice, J.\ Phys.\ A. {\bf 32}, 5233 (1999), doi:10.1088/0305-4470/32/28/304.

\bibitem{rhh91} S.\ Roux, A.\ Hansen and E.\ L.\ Hinrichsen, A direct mapping between Eden growth model and directed polymers in 
random media, J.\ Phys.\ A: Math.\ Gen.\ {\bf 24}, L295 (1991), doi.org/10.1088/0305-4470/24/6/008.

\bibitem{hsbkgv18} A.\ Hansen, S.\ Sinha, D.\ Bedeaux, S.\ Kjelstrup, M.\ Aa.\ Gjennestad and M.\ Vassvik, 
Relations between seepage velocities in immiscible, incompressible two-phase flow in porous media, Transp.\ Por.\ Med.\
{\bf 125}, 565-–587 (2018), doi.org/10.1007/s11242-018-1139-6. 

\bibitem{s53} A.\ E.\ Scheidegger, Theoretical models of porous matter,
Producers Monthly, August, 17 (1953).

\bibitem{s74} A.\ E.\ Scheidegger, The physics of flow through porous media,
University of Toronto Press, Toronto, 1974.

\bibitem{rhs19} S.\ Roy, A.\ Hansen and S.\ Sinha, Effective rheology of two-phase flow in a capillary fiber bundle model,
arXiv:1902.07577.

\bibitem{dl76} G.\ Duvaut and J.\ L.\ Lions, Inequalities in mechanics and physics, volume 219 of Grundlehren der mathematischen Wissenschaften, Springer 1976.

\bibitem{gl89} R.\ Glowinski, and P.\ Le Tallec, Augmented Lagrangian and operator-splitting methods in nonlinear mechanics, 
Vol. 9. SIAM, 1989.

\bibitem{grhbtc90} E.\ Guyon, S.\ Roux, A.\ Hansen, D.\ Bideau, J.\ P.\ Troadec and H.\ Crapo, Non-local and non-linear problems in 
the mechanics of disordered systems: application to granular media and rigidity problems, Rep.\ Prog.\ Phys.\ {\bf 53}, 373 (1990),
doi.org/10.1088/0034-4885/53/4/001.

\bibitem{tb13} L.\ Talon and D.\ Bauer, On the determination of a generalized Darcy equation for yield-stress fluid in porous media 
using a Lattice-Boltzmann TRT scheme,  Eur.\ Phys.\ J.\ E, {\bf 36}, 139 (2013), doi.org/10.1140/epje/i2013-13139-3.

\bibitem{kz87} M.\ Kardar and Y.\ C.\ Zhang, Scaling of directed polymers in random media, Phys.\ Rev.\ Lett.\ {\bf 58}, 2087 (1987),
doi.org/10.1103/PhysRevLett.58.2087.

\bibitem{taph13} L.\ Talon, H.\ Auradou, M.\ Pessel and A.\ Hansen, Geometry of optimal path hierarchies, EPL, {\bf 103}, 30003 (2013),
doi.org/10.1209/0295-5075/103/30003.

\bibitem{wbhh96} K.\ E.\ Wennberg, G.\ G.\ Batrouni, A.\ Hansen and P.\ Horsrud, Band formation in deposition of fines in porous media,
Transp.\ Porous Med.\ {\bf 25}, 247–-273 (1996), doi.org/10.1007/BF00140983.

\bibitem{rh87} S.\ Roux and H.\ J.\ Herrmann, Disorder-induced nonlinear conductivity, EPL, {\bf 4}, 1227 (1987), 
doi.org/10.1209/0295-5075/4/11/003.

\end{thebibliography}
\end{document}